\documentclass[twocolumn,showpacs,preprintnumbers,amsmath,amssymb]{revtex4}

\usepackage{graphicx}
\usepackage{dcolumn}
\usepackage{bm}

\begin{document}

\title{How Amp\`ere could have derived the Lorentz Transformations}
\author{G. Faraco}
\email{gefa@mat.unical.it} \affiliation{Istituto Nazionale Fisica
Nucleare, Italy; Dipartimento di Matematica, Universit\` a della
Calabria, via P. Bucci 30b, 87036, Rende (Italy)}

\author{G. Nistic\`o}
\email{gnistico@unical.it}
\affiliation{Dipartimento di
Matematica, Universit\` a della Calabria, via P. Bucci 30b, 87036,
Rende (Italy)} \affiliation{Istituto Nazionale Fisica Nucleare,
Italy; Dipartimento di Matematica, Universit\` a della Calabria,
via P. Bucci 30b, 87036, Rende (Italy)}

\date{\today}

\begin{abstract}
Lorentz Transformations of Special Relativity are derived from two
postulates: the first is the Principle of Relativity, while the
postulate of invariance of the velocity of light, used in usual
derivations, is replaced by a law of electro-magneto-statics and
invariance of electrical charge. Our derivation does not require
the assumption of regularity conditions of the transformations,
such as linearity and continuity required by other derivations.
The level of the needed mathematics and physical concepts makes
the proposed derivation suitable for Secondary School.

\end{abstract}

 \maketitle

\section{Introduction}
In Special Relativity, Lorentz Transformations are usually derived
from two postulates: the Principle of Relativity stating that
``The laws of physics are the same in all inertial frames'', and
the law of the invariance of velocity of light (Einstein's second
postulate) \cite{Einstein}. The literature shows that Lorentz
Transformations can be derived without this second postulate; a
partial review can be found in \cite{Field}.

In his approach to ``relativity without light'', Mermin
\cite{Mermin} derives the relativistic addition law for parallel
velocities directly from the Principle of Relativity and ``a few
simple assumptions of smoothness and symmetry''.  Starting from
the same assumptions of Mermin, Singh \cite{Singh} derives the
Lorentz Transformations.  The pedagogical advantage of these
approaches is that the invariance of light velocity, with its
paradoxical character, is not imposed \textit{ab initio}, but it
is obtained as a consequence of the theory. However, Mermin
him-self acknowledges that his approach entails an ``higher level
of analysis'', hence it ``is unavailable for a general educational
physics course, but as an introduction to special relativity for
physics majors''.

Another approach is due to Sen \cite{Sen}. He was able to obtain
essentially the same results of Mermin by using only simple
algebra and some assumptions of regularity of the functions
involved in his derivations. For this reason this approach can be
used for introducing the Special Relativity to the students at an
introductory level.

We have to remark that all these derivations do not single out
Lorentz Transformations as the only transformations consistent
with their postulates. Actually, what can be stated is that the
function $w(u,v)$ expressing the additional law of velocities $u$
and $v$ is $w(u,v)=(u+v)/(1+Kuv)$, where $K$ is a non-negative
constant. Therefore we have a large range of possibilities: if $K$
is $0$ then we get the Galilean Transformations; if $K>0$ then we
get a theory in which there is an invariant velocity
$c_K=1/\sqrt{K}$. Thus, what is the right transformation law
cannot be decided on the basis of their assumptions only.

In this work we propose a derivation of Lorentz Transformations,
and hence of Special Relativity, without making use of invariance
of light velocity. Similarly to the approaches above outlined, we
assume the Principle of Relativity and a few simple assumptions of
symmetry and reciprocity; nevertheless, our derivation is
different. We impose that a given law of electro-magneto-statics,
whose empirical validity was known since Amp\`ere \cite{Ampere},
holds in all inertial frames according to the Principle of
Relativity. In so doing we can directly derive Lorentz Contraction
and thereby Lorentz Transformations in their complete form without
indeterminate parameters, as the only transformations consistent
with the validity of that physical law in all inertial frames. In
our derivation, furthermore, it is not necessary making use of
regularity assumptions such as linearity or continuity. For this
reason, our derivation lies on (empirical) physics rather than on
mathematics.

A central point of the proposed derivation is to single out this
particular law. It is obtained by reformulating the conditions for
equilibrium of the classical electro-magneto-statics without
resorting to the notion of force, in a particular ideal experimental
situation. The main contribution to discovery of this rule was
given by Amp\`ere, who established its empirical validity
independently of any underlying theory.

Section II is devoted to establish the above mentioned law,
denoted by \textbf{(L.3)}. Furthermore, it is proved that
Principle of Relativity implies that the lengths orthogonal to be
relative motion between inertial frames are invariant (II.A). In
section III.A we derive the relation between lengths parallel to
the direction of the relative motion (Lorentz Contraction), by
using law \textbf{(L.3)}. Lorentz Transformations, in their
explicit form, are derived in section \textbf{(III.B)}, from
Lorentz Contraction. Our conclusive remarks are given in section
IV.

\section{A law of electro-magneto-statics}

In our derivation of Lorentz Transformations we need to impose
that, according to Principle of Relativity, a same law holds in two
different reference frames. In the present section, first we show
that Principle of Relativity implies that lengths orthogonal to
the direction of the relative motion between two inertial frames
are invariant; second we assume that electrical charge is
invariant, and finally establish a law of electro-magneto-statics
submitted to Principle of Relativity and  which can be expressed
without the concept of force.

\subsection{Invariance of transversal lengths}

\begin{figure}[htb!]
\centerline{\includegraphics[width=6cm]{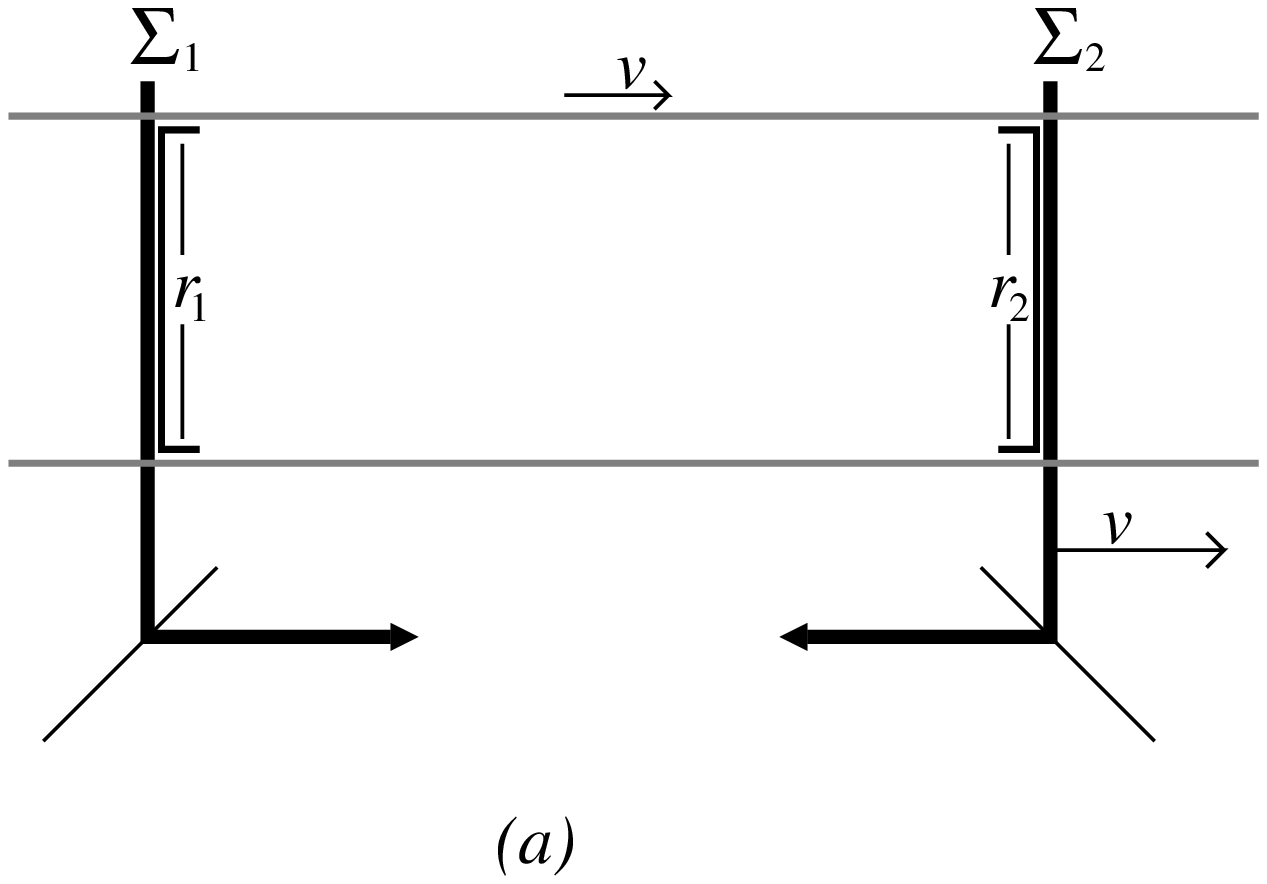}}
\caption{\label{fig1}}
\end{figure}

Let us consider two inertial frames $\Sigma_1$ and $\Sigma_2$,
with the $x$ axes oriented as in Fig. \ref{fig1} and with relative
constant velocity ${\bf v}=(v,0,0)$. Let us suppose that two
material wires are parallel to the direction of the relative
motion, and that one wire is at rest with respect to $\Sigma_1$,
while the other is at rest with respect to $\Sigma_2$. By $r_1$
and $r_2$ we denote the distances between the two wires with
respect to $\Sigma_1$ and $\Sigma_2$, respectively. Between these
distances, which are the measures of lengths orthogonal to the
direction of the relative motion,

one of the following two relations must hold:
$$
(i)\; r_1\leq r_2\qquad\hbox{or}\qquad (ii)\; r_2\leq r_1.
$$
If case (i) were realized, the following statement should hold as a physical law.
\begin{description}
\item[({\bf L.1})] {\sl Let $\Sigma_1$ be an inertial frame. If $\Sigma_2$ moves with
velocity ${\bf v}=(v,0,0)$ with respect to $\Sigma_1$,
the distances $r_1$ and $r_2$, relative to $\Sigma_1$ and $\Sigma_2$,
between two wires parallel to the direction of the relative motion, one at rest with respect
to $\Sigma_1$ and the other at rest with respect to $\Sigma_2$,
satisfy the following relation.
$$
r_1\leq r_2
.
$$}
\end{description}
The Principle of Relativity
implies that ({\bf L.1}) holds in every inertial frame $\Sigma_1$.
\par
\begin{figure}[htb!]
\centerline{\includegraphics[width=6cm]{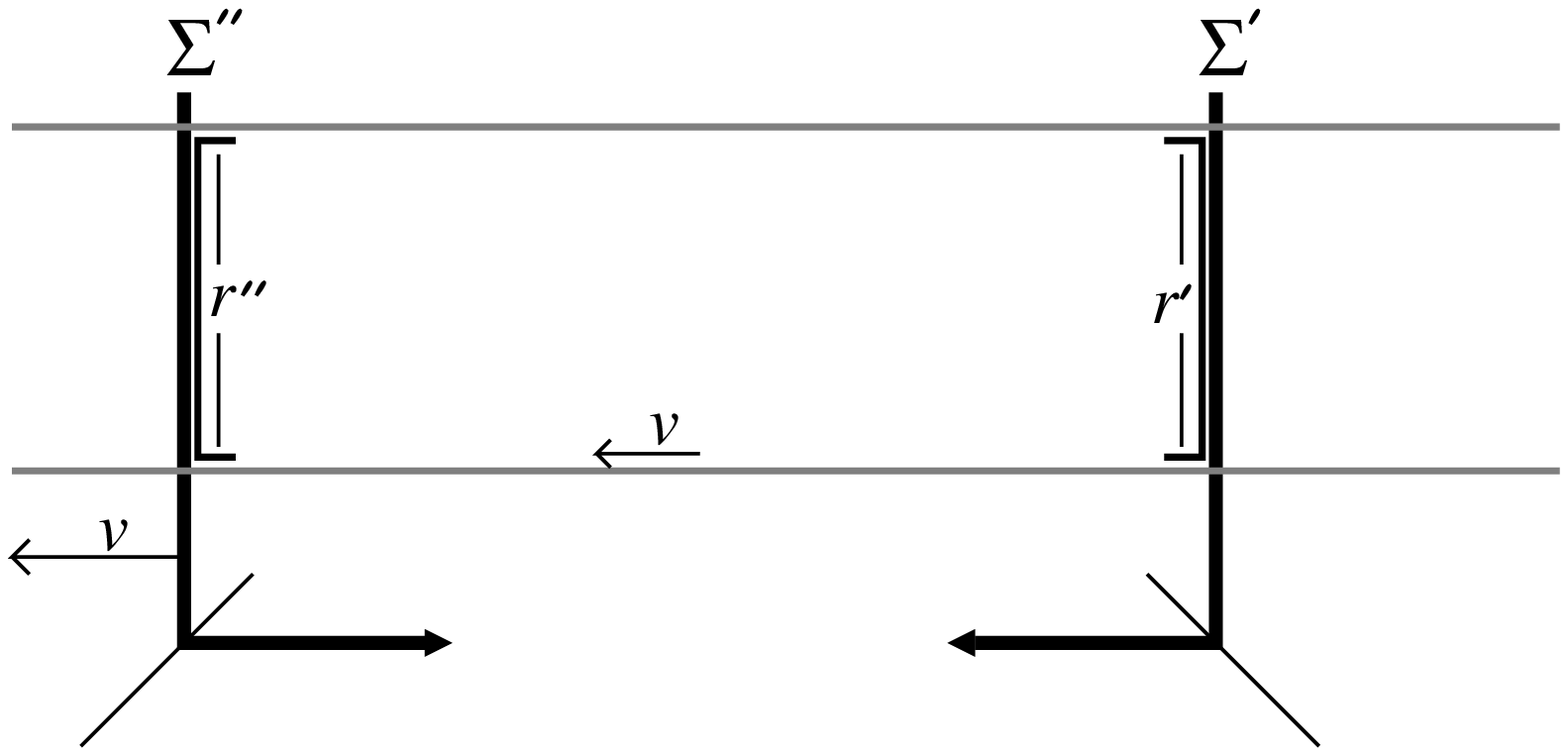}} \caption[]{}
\label{fig2}
\end{figure}

Now, let us denote frames $\Sigma_1$ by $\Sigma^{\prime\prime}$,
frame $\Sigma_2$ by $\Sigma'$, distance $r_2$ by $r'$ and distance
$r_1$ by $r^{\prime\prime}$. After this re-wording we see that
$\Sigma^{\prime\prime}$, $r'$ and $r^{\prime\prime}$, with respect
to $\Sigma'$, satisfy the conditions that ({\bf L.1}) requires to
$\Sigma_2$, $r_1$ and $r_2$ with respect to $\Sigma_1$. In
particular, one of the wires is at rest with respect to $\Sigma'$
(namely, the wire at rest with respect to $\Sigma_2$) and the
other is at rest with respect to $\Sigma^{\prime\prime}$ (Fig.
\ref{fig2}). Therefore law ({\bf L.1}) applies, and we have to
conclude that $r'\leq r^{\prime\prime}$; hence, $r_2\leq r_1$
holds together with $r_1\leq r_2$, so that $r_1=r_2$.

If case (ii) were realized, we would reach the same conclusion, by
using the same argument.

\subsection{Invariance of electrical charge}
Electrical charge is assumed to be invariant.

The quite familiar experience that if a solid body, with no (net) charge
is heated, then its (net) charge remains zero provides an argument showing invariance of charge.
Indeed, the increase of the
temperature provokes an increase of the average kinematic energy of
the particles constituting the body, i.e. electrons and nuclei.
But due to the lower mass of electrons, their velocity increases
much more than the velocity of nuclei. If the charge were
dependent on the velocity, the change of charge due to electrons
should overcome the change due to nuclei, and the body would
acquire a (net) electrical charge. But this phenomenon has never
been observed.

\subsection{The empirical law}

Let us consider
two parallel
wires which carry a stationary electrical current $i$ and an uniform
charge density $\lambda$. One of the laws of Electro-magneto-statics establishes that
\begin{description}
    \item[({\bf l.2})] The function
\begin{equation}\label{l.2}
\phi(\lambda, i)=\frac{\lambda^2}{2\pi\epsilon_0
r}-\frac{\mu_0}{2\pi}\frac{i^2}{r}
\end{equation}
is the density of the {\sl force} acting on each wire.
\end{description}
Unfortunately, this physical law can be interpreted only by making
resort to the notion of Newtonian {\sl force}. Since the effects
of a Newtonian force follow from the second Newton law
\par
\textbf{(NL)} \qquad\qquad{\sl force }= {\sl mass} $\times$ {\sl
acceleration}
\par\noindent
the validity of \textbf{(l.2)} could be verified only by
implicitly assuming \textbf{(NL)}. To avoid such a further
undesired assumption, we shall extract from \textbf{(l.2)} another
physical law, with a more poor physical content than
\textbf{(l.2)}, but which can be expressed without using undesired
concepts, such as that of force, and whose empirical validity can
be tested independently of any theory.
\par
Let us consider a device $\cal D$ consisting of an uniform
distribution of identical springs, each of them acting on both
wires. If the springs are suitably manufactured, the action of the
device establishes equilibrium between the wires. An implication
of \textbf{(l.2)} is that such an equilibrium is broken if
$\lambda$ and $i$ change their values in such a way that also the
value of $\phi$ turns out to be modified.
This statement can be re-formulated as follows.
\begin{description}
    \item[({\bf L.2})]
If the action of device $\cal D$ yields equilibrium for both the
two pair of values $(\lambda,i)$ and $(\hat{\lambda},\hat{i})$,
then
$$
\phi(\hat{\lambda}, \hat{i})=\phi(\lambda,i)
$$
\end{description}
Statement ({\bf L.2}) is an empirical implication of the laws
of electromagnetostatics, which is expressed without the need to
interpret function $\phi$ in terms of {\sl force}. Principle of
Relativity implies that ({\bf L.2}) must be considered as a
physical law which holds in all inertial frames.
\\
{\bf Remark 1.} The equilibrium could be obtained by means of a
device different from $\cal D$, for instance by replacing the
springs with a greater number of weaker ones, without affecting
the validity of ({\bf L.2}). Now we show that the device can be
designed in such a way to be {\sl invariant}.

\begin{figure}[htb!]
\centerline{\includegraphics[width=6cm]{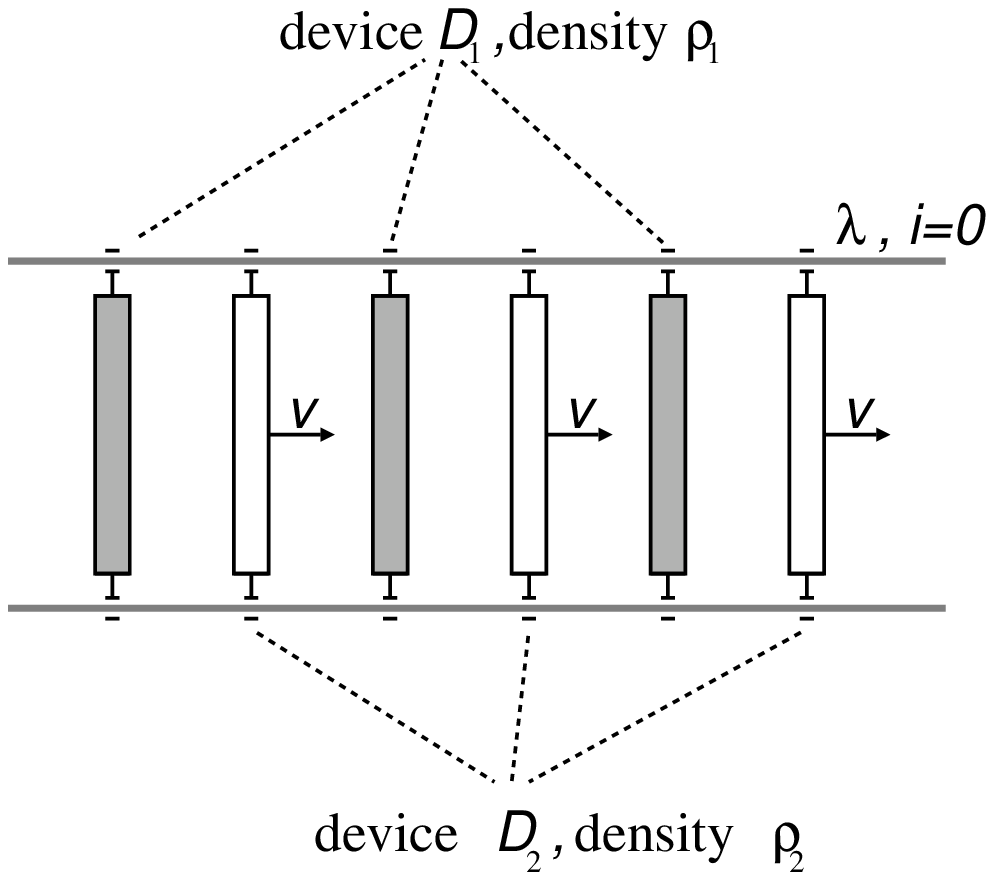}}
\caption[]{Device ${\cal I}={\cal D}_1+{\cal D}_2$ with respect to
$\Sigma$} \label{fig3}
\end{figure}

Given the two wires considered above in frame $\Sigma$, let us
consider another frame $\Sigma'$ which moves with respect to frame
$\Sigma$ with a constant velocity $v$ in the direction parallel to
the wires (Fig. \ref{fig3}). Our device $\cal I$ consists of two
distributions of springs, ${\cal D}_1$ and ${\cal D}_2$. The first
one, ${\cal D}_1$, is made up of identical springs, each of them
acting on both wires, uniformly distributed with density $\rho_1$,
at rest with respect to frame $\Sigma$. Second distribution,
${\cal D}_2$, is made up of springs which move with velocity $v$
(hence they are at rest with respect to $\Sigma'$), identical to
each other, but different from the spring of ${\cal D}_1$,
uniformly distributed with density $\rho_2$ with respect to
$\Sigma$. Hence, with respect to $\Sigma$ device $\cal I$ consists
of a distribution at rest with density $\rho_1$ and another
distribution with density $\rho_2$ which moves with velocity $v$.

\begin{figure}[htb!]
\centerline{\includegraphics[width=6cm]{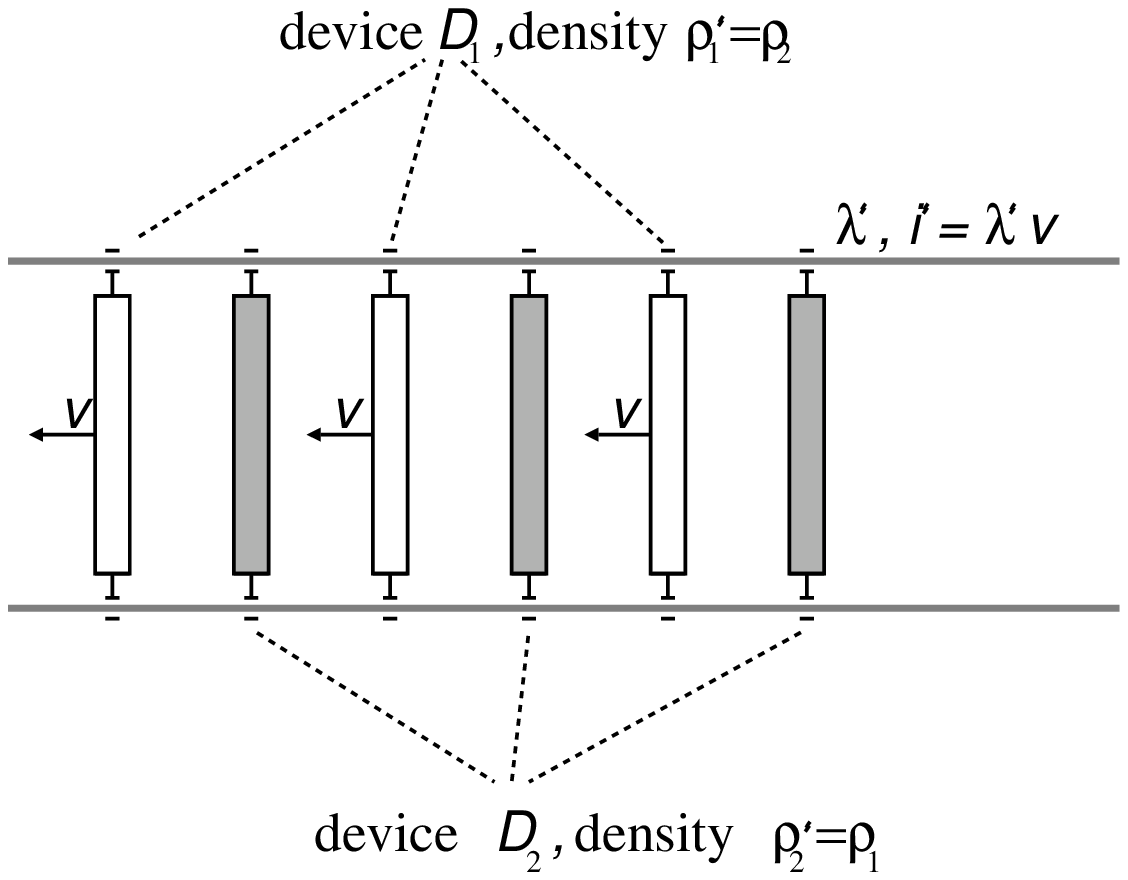}}
\caption[]{Device ${\cal I}={\cal D}_1+{\cal D}_2$ with respect to
$\Sigma'$} \label{fig4}
\end{figure}

Now, the value $\rho_2$ of the density of ${\cal D}_2$ with
respect to $\Sigma$ is chosen in such a way that its value
$\rho_2'$ with respect to $\Sigma'$ turns out to be equal to
$\rho_1$. This means that if a distribution at rest has density
$\rho_1$, then it has density $\rho_2$ with respect to a frame
where it moves with velocity ${v}$.
As a consequence, with respect to $\Sigma'$,
device $\cal I$ consists of a distribution at rest with density
$\rho_1$ and another distribution with density $\rho_2$ and
velocity ${v}$ (Fig. \ref{fig4}).
\par
Therefore, as regards to the densities, device $\cal I$ composed
by ${\cal D}_1$ and ${\cal D}_2$ appears to $\Sigma$ identical to
that seen by $\Sigma'$, apart from an exchange of the roles of
${\cal D}_1$ and ${\cal D}_2$. Like the density $\rho_2$, the
value of any other physical magnitude of ${\cal D}_2$, which
determines its action on the wires, can be chosen in such a way
that device ${\cal I}={\cal D}_1+{\cal D}_2$ appears to $\Sigma'$
identical to that seen by $\Sigma$, after an exchange in the roles
of ${\cal D}_1$ and ${\cal D}_2$. For instance, suppose that the
value of one of these magnitudes, say $C$, is $c_1$ for the
springs of ${\cal D}_1$ with respect to $\Sigma$. Then we choose
the springs of ${\cal D}_2$ with a value $c_2$ of $C$ with respect
to $\Sigma$ such that its value $c_2'$ with respect to $\Sigma'$
satisfies $c_2'=c_1$, which implies $c_1'=c_2$ (in fig.4 value
$c_1$ (resp., $c_2$) is revealed by the white (resp. gray) colour
of the spring). The invariance of $\cal I$ is completed by the
fact that the distance $r$ between the wires is invariant, as
proved in section II.A
\par
Now, according to \textbf{remark 1}, we state the following law
({\bf L.3}), obtained from ({\bf L.2}) by replacing device $\cal
D$ by the invariant device $\cal I$.
\begin{description}
    \item[({\bf L.3})]
If the action of device $\cal I$ yields equilibrium for both the
two pair of values $(\lambda,i)$ and $(\hat{\lambda},\hat{i})$,
then
$$
\phi(\hat{\lambda}, \hat{i})=\phi(\lambda,i).
$$
\end{description}

\section{Lorentz transformations}

In this section we shall establish which are the transformation
laws consistent with the Principle of Relativity and law {(\bf L.3)}.
To do this, we derive a relation between the lengths observed by
two different frames of reference. While the lengths orthogonal to
the direction of relative motion are invariant, in III.A we show
that the lengths parallel to the direction of relative motion are
not invariant.

\subsection{Lorentz contraction}

We consider two wires with uniform electrical charge, at rest in
$\Sigma$. Let us introduce another reference frame $\Sigma'$ that
moves with respect $\Sigma$ with velocity ${v}$ in the same
direction of the wires.  For realizing the equilibrium we consider
a device $\cal I$, like that designed in sect. II.C, manufactured
in such a way that it is invariant. Since the equilibrium is
realized in both $\Sigma$ and $\Sigma'$ by means the same device,
by {\bf (L.3)} we have to conclude that
\begin{equation}
\label{A} \phi(\lambda, i)=\phi(\lambda', i')
\end{equation}
where $\lambda$ ($i$) and $\lambda'$ ($i'$) are the values of
density charge (current) with respect $\Sigma$ and $\Sigma'$
respectively. Since the wires in $\Sigma$ are at rest, we have $i=0$. In $\Sigma'$ the current is produced by the motion
of the wires, therefore $i'=\lambda'{v}$. Then (\ref{A})
becomes
$$\frac{\lambda^2}{2\pi\epsilon_0
r}=\frac{\lambda'^2}{2\pi\epsilon_0
r}-\frac{\mu_0}{2\pi}\frac{(\lambda'v)^2}{r}$$
 which implies
$\lambda^2=\lambda'^2(1-{\epsilon_0\mu_0}{v^2})$; therefore,
if we set $\epsilon_0\mu_0=\frac{1}{c^2}$, we have
\begin{equation}\label{C}
\lambda=\lambda'\sqrt{1-\frac{v^2}{c^2}}
\end{equation}
This result says that the charge density is not invariant.
\par
Now we consider a piece of wire of length $L$ carrying a charge
$\delta Q$ respect to $\Sigma$. With respect to $\Sigma'$, this same piece of wire has a length $L'$ and carries a
charge $\delta Q'=\delta Q$. Then
$$\lambda=\frac{\delta Q}{L}\quad\quad \textrm{and}\quad\quad
\lambda'=\frac{\delta Q'}{L'}=\frac{\delta Q}{L'}.$$
Therefore, (\ref{C}) becomes
$$
\frac{\delta Q}{L}=\frac{\delta
Q}{L'}\sqrt{1-\frac{v^2}{c^2}}
$$
which leads to
\begin{equation}\label{B}
L'=L\sqrt{1-\frac{v^2}{c^2}}
\end{equation}
This relation is known as \textsl{Contraction of Lorentz}.

\subsection{From Lorentz contraction to Lorentz transformations}

Now we derive Lorentz Transformations from Lorentz Contraction. Our aim is to obtain the law of motion
of a particle $P$ with respect to $\Sigma'$ when this law is known
in $\Sigma$. We proceed as follows:
\begin{description}
    \item[(a)] first, we consider the case in which particle $P$
    is at rest in $\Sigma$ on the $x$ axis and we use Lorentz
    Contraction to derive its law of motion in $\Sigma'$;
    \item[(b)] the result obtained in (a) is generalized to a particle at rest in a any spatial point of $\Sigma$.
    \item[(c)] by means of the results of (b), we derive the law of
    motion in $\Sigma'$ when the particle moves in
    $\Sigma$ according to any known law.
\end{description}
{\bf Step (a)}.  If  particle $P$ is at rest in the point of
coordinate $\xi$ of the $x$ axis with respect to $\Sigma$, its
motion with respect to $\Sigma'$ is described by the ``world
line'' $(\tau',\xi'(\tau'))$, where $\xi'(\tau')$ is the $x$
coordinate of the particle at time $\tau'$ with respect to
$\Sigma'$. The particle moves with respect to $\Sigma'$ with a
velocity $-{v}$.

The value $\xi$ represents the length $l$ of the segment $[0,\xi]$
on the spatial $x$ axis of $\Sigma$. This length $l$ is related to
the length $l'$ of this same segment with respect to $\Sigma'$
just by Lorentz Contraction
\begin{equation}\label{E}
l'=l\sqrt{1-{v^2\over c^2}}.
\end{equation}
But $l'$ is also the difference between the coordinates of $P$ and
of the origin of $\Sigma$ with respect to $\Sigma'$, which are
$\xi'(\tau')$ and $-v\tau'$:
\begin{equation}\label{D}
l'=\xi'(\tau')-(-v\tau')=\xi'(\tau')+v\tau',
\end{equation}
Therefore, by equating (\ref{E}) and (\ref{D}) we get
\begin{equation}\label{F}
\xi'(\tau')=\xi\sqrt{1-{v^2\over c^2}}-v\tau'.
\end{equation}
This relation is the law of motion with respect to $\Sigma'$ in
the case (a). The same argument can be used to show that if a
particle is at rest in point $\xi'$ with respect to $\Sigma'$,
then its world line $(\tau, \xi(\tau))$ with respect to $\Sigma$
is given by
\begin{equation}\label{G}
 \xi(\tau)= \xi'\sqrt{1-{v^2\over
c^2}}+v\tau.
\end{equation}
{\bf Step (b).} Now we consider a particle $P$ at rest in the
point of coordinates $(\xi, \eta, \zeta)$ with respect to
$\Sigma$. Our aim is to find the world line $(\tau', \xi'(\tau'),
\eta'(\tau'), \zeta'(\tau'))$ with respect to $\Sigma'$. Let us
imagine a parallelepiped at rest in $\Sigma$ with a vertex in the
origin of $\Sigma$ three edges lying along the axes $(x, y, z)$,
and particle $P$ in  the vertex with the greatest distance from
the origin. With respect to $\Sigma'$ the coordinates
$(\xi(\tau'), \eta'(\tau'), \zeta'(\tau'))$ of $P$ at time $\tau'$
coincide with those of this last vertex. The coordinates
$\eta'(\tau')$ e $\zeta'(\tau')$ are the lengths of the edges of
the parallelepiped orthogonal to the relative motion, and
therefore are invariant. For the $x$ coordinate, we can repeat the
argument of step (a); thus, with respect to $\Sigma'$, particle
$P$ moves according to
\begin{equation}\label{H}
  \begin{cases}
     \xi'(\tau')=\xi\sqrt{1-{v^2\over c^2}}-v\tau'& \text{}, \\
    \eta(\tau')=\eta & \text{} \\
\zeta(\tau')=\zeta & \text{}
  \end{cases}
\end{equation}

Reciprocally if a particle is at rest in  $\Sigma'$ its laws
motion with respect to $\Sigma$ are
\begin{equation}\label{I}
  \begin{cases}
    \xi(\tau)= \xi'\sqrt{1-{v^2\over c^2}}+v\tau & \text{}, \\
     \eta(\tau)=\eta(\tau') & \text{} \\
\zeta(\tau)=\zeta(\tau')& \text{}
  \end{cases}
\end{equation}
{\bf Step (c).}  Now we let particle $P$ move in an arbitrary way
with respect to $\Sigma$; suppose  that at time $t$ it is in the
point $(x, y, z)$. Let us imagine a particles $A$ at rest with
respect to $\Sigma$ and another particles $B$ at rest in
$\Sigma'$, such that both $A$ and $B$ collide with our particle
just in the space-time point $(t, x, y, z)$. Our assumption  is
simply that this threefold collision occurs also in $\Sigma'$ in a
space-time point denoted by $(t',x', y', z')$. Since $B$ is at
rest in $\Sigma'$ its position is just the location of the impact,
i.e. $(x', y', z')$. The space-time point of the collision  must
belong to the world line of particle $A$ in $\Sigma'$. Therefore
by (\ref{H}) we have
\begin{equation}\label{L}
  \begin{cases}
     x'=x\sqrt{1-{v^2\over c^2}}-vt' & \text{}, \\
    y'=y & \text{} \\
z'=z & \text{}
  \end{cases}
\end{equation}
and by (\ref{I})
\begin{equation}\label{M}
  \begin{cases}
    x=x'\sqrt{1-{v^2\over c^2}}+vt & \text{}, \\
    y'=y & \text{} \\
z'=z & \text{}
  \end{cases}
\end{equation}
By rewriting (\ref{L}) and (\ref{M}) in explicit form, we get the
usual form of Lorentz Transformations:
\begin{equation}\label{N}
  \begin{cases}
    x'=\frac{x-{v}t}{\sqrt{1-\frac{v^2}{c^2}}} & \text{}, \\
    y'=y & \text{} \\
z'=z & \text{} \\
t'=\frac{t-\frac{{v}}{c^2}x}{\sqrt{1-\frac{v^2}{c^2.}}} & \text{}
  \end{cases}
\end{equation}

\section{Conclusions}

The heart of the present derivation is  law \textbf{(L.3)}. Since
this rule is obtained from \textbf{(l.2)}, which has not an
interpretation independent of the concept of force, the legitimacy
of its use could be questioned. However, these kinds of doubts are
ruled out by the following arguments. Once function $\phi$ in
\textbf{(L.3)} is interpreted as empirical mean to establish
conditions for the equilibrium, its empirical validity can be
experimentally verified without making reference to any underlying
theory. This is the method followed by Amp\`ere in establishing
the ``Th\'eorie des ph\'enom\`enes \'electro-dynamiques,
uniquement d\'eduite de l'exp\'erience'' \cite{Ampere}.

Amp\`ere him-self wrote: ``\textsl{The main advantage of the
formulas so established [...] is that of remaining independent of
the hypotheses, both of them used by their authors in the research
of the formulas, and also of the hypotheses that in the future
replace the former.  [...] Whatever the physical cause one wants
attribute to the phenomena made by such an [electro-dynamical]
action, the formula obtained by my-self will be always the
expression of real facts. [...] The adopted [method] which led me
to the desired results [...] consists in verifying, by means of
experience, that an electrical conductor rests in equilibrium
under equal forces [...]}''

Clarified the legitimacy of the arguments presented in our
derivation of Lorentz Transformations, we would now to emphasize
the advantage that our approach carry by a didactics and
pedagogical point of view. The derivations of Lorentz
Transformations without light produce that the second postulate of
Einstein  is as a direct consequence of the theory and, for this
reason, it can be easily accepted by the student that frequently
reject concepts that cannot be observed and verified
experimentally. In addition, the proposed approach allows to
obtain the Lorentz Transformations as the only transformations
consistent with the starting assumptions. Other derivations without
light, on the contrary, exhibit that there are only two possible
equations of transformations, one corresponding to the old
Galilean-Newtonian transformation laws and the other corresponding
to the standard Lorentz ones without specifying the value of the
velocity of light. The starting assumptions are very few, do not
concerning linearity and continuity of the functions of
transformations; the linearity and continuity are properties that
the student can directly verify. The use of physical concepts and
laws already known by students enables teachers to use this
approach in a introductory physics class as well, and furthermore
in a Secondary School in which the Lorentz Transformation are
presented, but not derived, as new laws of transformations to be
substituted to the Galilean Transformation in order to take into
account the time dilatation and length contractions and other
effects which follow from the Relativity Theory.

\bibliography{apssamp}

\end{document}